\documentclass[12pt,times,draftclsnofoot,a4paper]{elsarticle}

\usepackage{graphicx}
\usepackage{lineno}
\usepackage{xspace}
\usepackage{lipsum}
\usepackage{cleveref}
\usepackage{stackengine}
\usepackage{titlesec}
\usepackage{xcolor}
\usepackage{comment}
\usepackage[bitstream-charter]{mathdesign}
\usepackage[T1]{fontenc}
\usepackage{geometry, float}
\usepackage[normalem]{ulem}
\usepackage{stackengine}

\journal{Nuclear Instruments and Methods in Physics Research A}

\begin{document}

\begin{frontmatter}

\setcounter{page}{0}
\title{Simulation of angular resolution of a new electromagnetic sampling calorimeter}

\author[jbnu]{Junlee~Kim}
\ead{junlee.kim@cern.ch}

\author[jbnu]{Eun-Joo~Kim\corref{cor1}}
\ead{ejkim@jbnu.ac.kr}
\cortext[cor1]{Corresponding author. Tel.: +82-10-4581-5649.}

\author[korea]{Young~Jun~Kim}
\author[korea]{Jung~Keun~Ahn}
\author[kek]{Gei~Youb~Lim}

\address[jbnu]{Division of Science Education, Jeonbuk National University, Jeonju 54896, Korea}
\address[korea]{Department of Physics, Korea University, Seoul 02841, Korea}
\address[kek]{Institute of Particle and Nuclear Studies (IPNS), High Energy Accelerator Research Organization (KEK), Tsukuba 305-0801, Japan}


\begin{abstract}
We report on the simulation results for the angular resolution of an electromagnetic (EM) sampling calorimeter with photons in the range of 100~MeV to 2~GeV. The simulation model of the EM calorimeter consists of alternating layers of a 1-mm-thick lead plate and a 5-mm-thick plastic scintillator plate. The scintillator plates are alternately segmented into horizontal and vertical strips. In this study, we obtain energy deposits in individual strips using Geant4 simulations and reconstruct the incident photon angles using XGBoost with gradient-boosted decision trees. The performance of the angle reconstruction depends on the detector configuration and the accuracy of machine learning. The angular resolution is well described by the expression $0.24^{\circ} \oplus 1.25^{\circ}/\sqrt{E_{\gamma}}$, where $E_{\gamma}$ is the incident photon energy in GeV, for strips of 15 mm and 32 layers. This energy dependence is consistent for different incident angles in the range of 10$^{\circ}$ to 40$^{\circ}$.

\end{abstract}
\begin{keyword}
Electromagnetic Calorimeter \sep Geant4 \sep XGBoost
\end{keyword}

\end{frontmatter}

\section{Motivation}
\label{sec:mot}

Calorimeter plays a crucial role in the experimental studies of nuclear and particle physics~\cite{unitext}. An ideal calorimeter fully absorbs the energy of an entering particle and converts the energy into measurable quantities. The sampling calorimeter consists of alternating layers of two different materials, one is an absorber generating a shower of secondary particles, and the other is an active medium that generates signals from the energy deposits of the secondaries. Because the energy deposit in the active medium fluctuates with respect to the energy loss in the dense absorber, the sampling calorimeter has an inevitable drawback in energy resolution, especially in the gamma measurement. However, the energy resolution can be improved by optimizing the alternating layer configuration.  In addition, it becomes less critical as gamma energy becomes higher.

In general, the sampling calorimeter enables cost-effective fabrication of large-scale detectors and design of flexible geometric configuration. This detector has been developed in various ways by selecting a material of absorber among many candidates such as brass~\cite{CMS:mat}, lead~\cite{CDF:mat}, tungsten~\cite{DELPHI:mat}, uranium~\cite{UGAS:mat}, and also of active medium such as scintillators~\cite{CMS:mat,CDF:mat}, gas~\cite{UGAS:mat}, solid-state detector~\cite{DELPHI:mat}, and liquid gas~\cite{LiqAR:mat}. Also, there are many topologies to pile up the materials and read out their signals, such as sandwich~\cite{KOTO:MB}, shashlik~\cite{shashlik:con}, spaghetti~\cite{KLOE:con}, and accordian~\cite{LiqAR:mat}.

The sampling calorietmer has been used to measure the direction of arriving photons in a space-based detector to identify astrophysical sources for gamma-ray astrophysics. The Fermi Large Area Telescope (LAT) has a tracker consisting of 18 $x$-$y$ pairs of silicon strip detectors with 16 interleaved tungsten foils. Tracking with the silicon detectors on the electron-positron pairs produced at the foils provides an angular resolution of 0.6 degrees for 1-GeV photons and better than 0.16 degrees for photons with energies higher than 10~GeV~\cite{FERMI:LAT}. The LAT has two additional components: a plastic scintillator for the charged-particle detection and a homogeneous CsI crystals for the energy measurement. This triple detector configuration is also used in different missions. An energy spectrum measurement of cosmic-ray electrons (CALET)~\cite{CALET} utilizes scintillating fibers and PWO crystals instead of the silicon strips and the CsI crystals, respectively. A dark matter particle explorer mission (DMPE)~\cite{DMPE} selects BGO crystals for the energy measurement. Theses detectors provide the angular resolution of 1-2 degrees for 1-GeV photons.

In the accelerator-based experiments, the direction of photons can be precisely determined by connecting the interaction position obtained from charged particle tracks to their incident position measured using the calorimeter. However, when observing $K_{L} \rightarrow \pi^{0}\nu\bar{\nu}$ decay in the absence of charged particles, it is impossible to reconstruct the interaction (decay) position. The decay occurs through the CP-violating Flavor Changing Neutral Current and has significant potential to provide clues to new physics beyond the Standard Model (SM). Its branching fraction is calculated to be extremely small as 3~$\times$~10$^{-11}$ in the SM~\cite{KOTOtheory}, which implies that the effect of new physics on the decay could overwhelm that of the SM. Any observed deviation from the SM prediction will indicate new physics without hesitation, thanks to the small uncertainty in the theoretical calculation.

The $K_{L} \rightarrow \pi^{0}\nu\bar{\nu}$ decay is identified as a final state existing only a single $\pi^{0}$ reconstructed from detected two photons. A hermetic detector system enclosing the decay region should be prepared to confirm only two photons in the final state. There had been an experimental proposal to detect the $K_{L} \rightarrow \pi^{0}\nu\bar{\nu}$ decay with the reconstruction of a single $\pi^{0}$ using directional information of two photons~\cite{KOPIO}. A Preradiator (PR) is placed in front of the calorimeter to measure the photon's incident angle. The PR consists of a close-packed multi-layer sandwich of 8-mm-thick plastic scintillators alternating with the single-plane and thin-gap drift chambers with cathode strips. For good energy resolution and high detection efficiency against a few hundred MeV photons, the plastic scintillator is used for electron-positron conversion. This idea had been confirmed by a beam test using tagged photon beams and achieved its angular resolution of 25~mrad for 250-MeV~photons~\cite{KOPIOcdr}.

On the other hand, the KOTO experiment at J-PARC~\cite{KOTOexp} takes another approach to identify $K_{L} \rightarrow \pi^{0}\nu\bar{\nu}$ decay with a well-collimated small-size neutral beam, which provides a condition that the $\pi^{0}$ decays at the beam axis. With this assumption, the decay vertex is calculated using the positions and energies of the two photons measured in the calorimeter. The transverse momentum of $\pi^{0}$ is calculated from the vertex. The two variables, the decay vertex and the transverse momentum are used to define a signal region of the decay. The KOTO experiment started data taking since 2013 and has achieved its experimental sensitivity as 7.2~$\times$~10$^{-10}$~\cite{KOTOexp}. The data taking will continue for a few more years, and the sensitivity will be expected to reach $\mathcal{O}$(10$^{-11}$).

A successive experiment of the KOTO is under preparation to observe more than 35 $K_{L} \rightarrow \pi^{0}\nu\bar{\nu}$ decays with a higher-intensity beam and a larger-scale detector system~\cite{KOTO2}. In this highly sensitive measurement, the halo kaon, which is scattered by beam line materials and decays at off-beam axis, will become a considerable background source when the kaon decays via $K_{L} \rightarrow 2\gamma$. In addition, the neutrons away from the beam axis produce a $\pi^{0}$ or $\eta$ by interacting with detector materials. Two photons from those events are also one of main backgrounds. We can remove them by comparing the incident angle expected from the reconstructed vertex and the actual measurement using a new sampling calorimeter in this paper.

In this respect, we report on results of Monte Carlo studies on directional measurement analyzing energy-weighted shower shapes generated in a toy sampling calorimeter, which consists of segmented alternating lead and plastic scintillator layers. There is a limitation in the angular resolution from the analysis of shower shape produced in a sampling calorimeter due to the stochastic feature of its generation. A deduction of the incident angle of the EM shower using a lead scintillating fiber sandwiched sampling calorimeter is studied~\cite{PbScint}, and the angular resolution is expected to be larger than 8 degrees for 1-GeV electrons. We take a machine-learning approach to get better angular resolution from the analysis. Recently, machine-learning analysis has become a critical tool for better measurement of particle identification~\cite{mlp}, timing~\cite{mlt}, and energy~\cite{mle}. We use the XGBoost (XGB), which is an optimized distributed gradient-boosting library~\cite{xgboost:2016}.

A description on the detector configuration and Monte Carlo generation will be given in Section~\ref{sec:ems}. In the Section~\ref{sec:res}, results of the machine learning studies on the generated data will be presented, and a summary will follow in the Section~\ref{sec:sum}. 

\section{Electromagnetic shower simulation}
\label{sec:ems}

\begin{figure}[!hbt]
\centering
\includegraphics[width=0.6\textwidth]{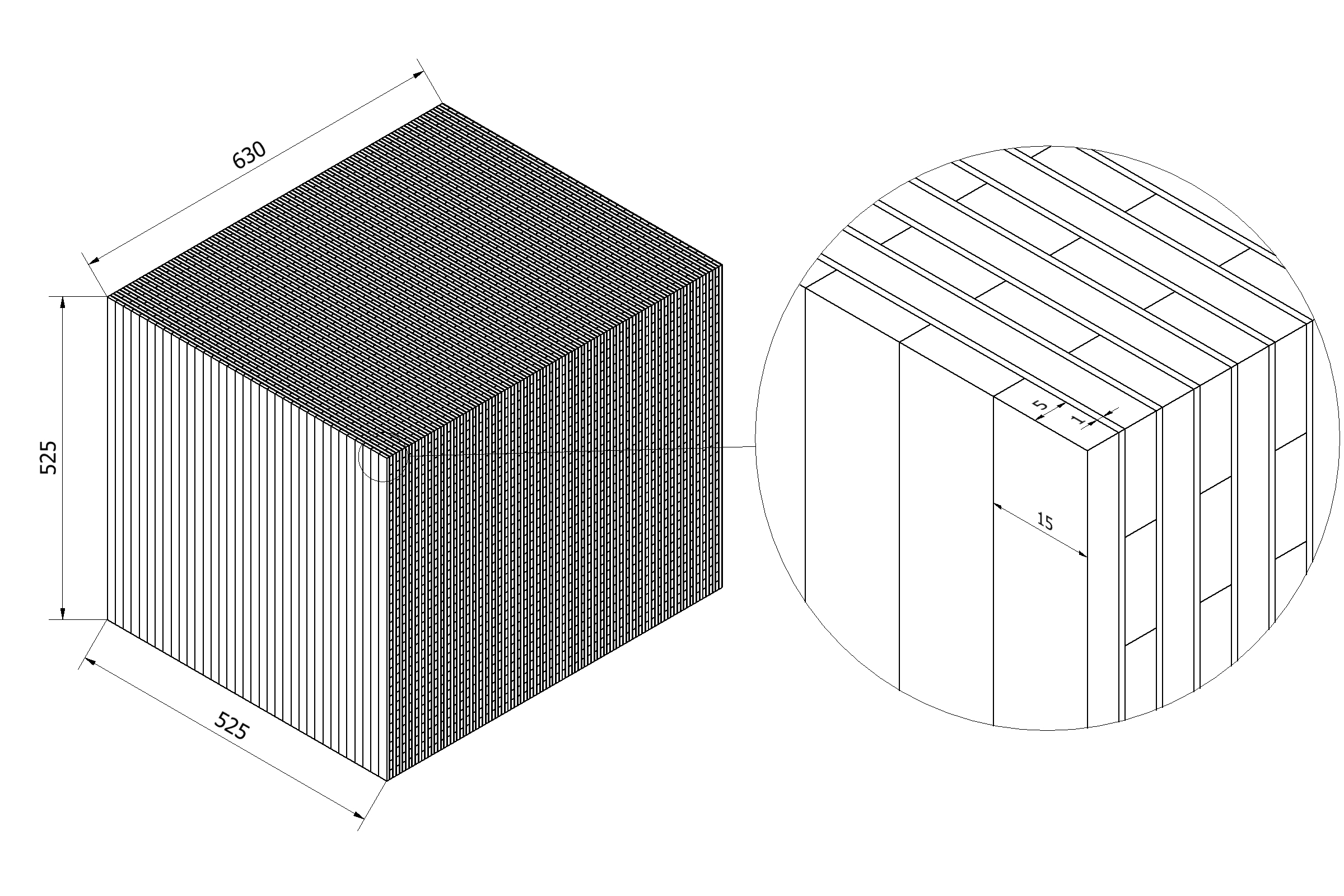}
\caption{ Schematic of the sampling calorimeter model consisting of 105 alternating layers of lead and scintillator plates. Each plate is segmented in 35 strips, oriented alternately in horizontal and vertical directions. }
\label{fig:det_conf}
\end{figure}

The toy sampling calorimeter is designed as a block consisting of alternating layers of an 1-mm-thick lead absorber and a 5-mm-thick polyvinyltoluene-based plastic scintillator. The plastic scintillator is segmented into 15-mm-wide strips, which are alternately oriented in the vertical and horizontal directions, as shown in Figure~\ref{fig:det_conf}. It has a cross-section size of 525~$\times$~525~mm$^{2}$ and accommodates 105 alternating layers of 630~mm and 20 radiation lengths (20$X_{0}$) to contain the EM shower of photons for energies in the range of 0.1 to 2~GeV.

\begin{figure}[!hbt]
\centering
\includegraphics[width=0.48\textwidth]{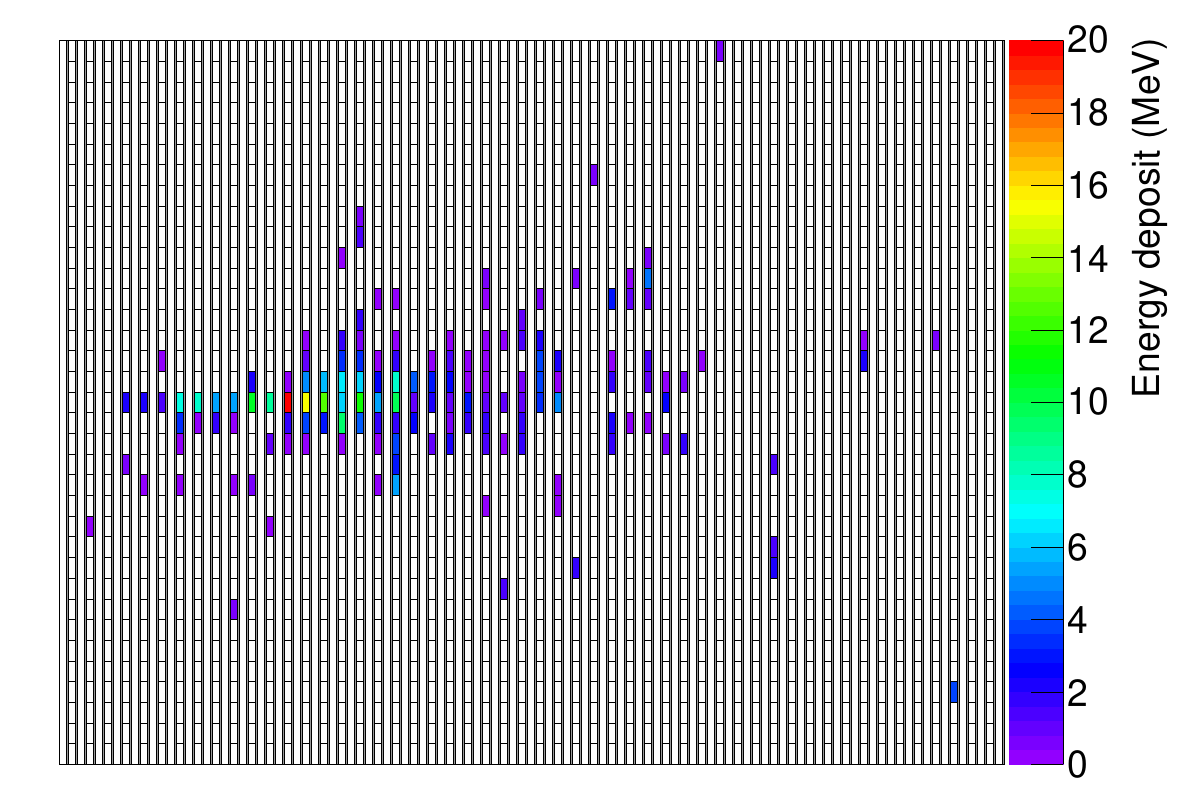}
\includegraphics[width=0.48\textwidth]{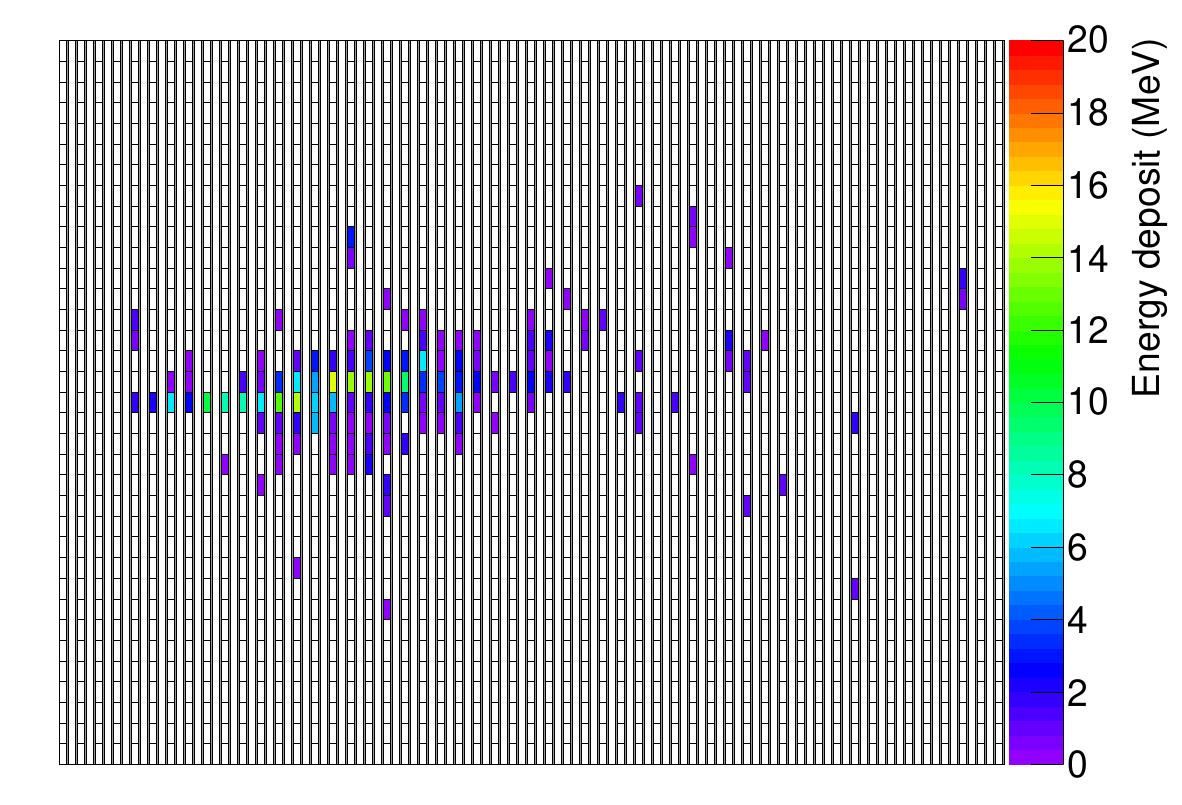}
\caption{ Event display of simulated energy deposit patterns for an 1-GeV photon entering the calorimeter ($\theta=$~0) in (a) $xz$- and (b) $yz$-planes.}
\label{fig:Evt_Dis}
\end{figure}

The detector response to the incident photons is simulated using Geant4 (ver. 4.10.06) with standard EM sub-packages~\cite{GEANT4}. The direction normal to the front surface defines the $z$-axis. The photon direction is defined by the polar angle ($\theta$) with respect to the $z$-axis. Figure~\ref{fig:Evt_Dis} illustrates the simulated energy deposit patterns in each strip for an 1-GeV photon at normal incidence in the $xz$- and $yz$-planes. Each segmented region shown in Figure~\ref{fig:Evt_Dis} represents a channel. 

Training data for machine learning are prepared using the Geant4 simulation such that the input dataset is representative of the detector response of the sampling calorimeter. In order to minimize the bias in the training process related to the dataset, the incident angles are uniformly generated at the detector surface in the range of 0~$<\theta<$~50$^{\circ}$ and 0~$<\varphi<$~360$^{\circ}$, where $\varphi$ denotes the azimuthal angle. The number of training samples is $10^{5}$ considering the limited computing resources. The angle reconstruction is studied using photons generated at given incident $\theta$ angle, energy, and position. It turns out unchanged with the energy deposit threshold above the noise level of 0.5 MeV. Therefore, we do not consider random noises in this study. 

\section{Results}
\label{sec:res}
\subsection{Reconstruction of incident angles}
\label{sec:reco}

The incident angle of the photons is reconstructed using XGB~\cite{xgboost:2016}, which maps a feature dataset, the energy deposit in each scintillator strip, onto a target variable, the incident angle of a photon. The feature size identical to the number of strips varies with the strip width to have the same coverage. In the case of the 15-mm-wide strips, 35~$\times$~105 channels compose the detector. XGB constructs gradient-boosted decision trees using the training dataset, efficiently descending the loss function. The hyperparameters of XGB drive how models are trained. In this paper, main five parameters are studied. \textit{N\_estimators} defines the maximum number of allowed decision trees to be developed, and \textit{Max. depth} defines the complexity of the decision-tree structure. \textit{Subsample} controls the fraction of total event samples for each boosting procedure, \textit{learning rate} weights a decision tree to be added onto the current model, and \textit{gamma} regulates the evaluation of each decision tree.

\begin{figure}[!hbt]
\centering
\includegraphics[width=0.7\textwidth]{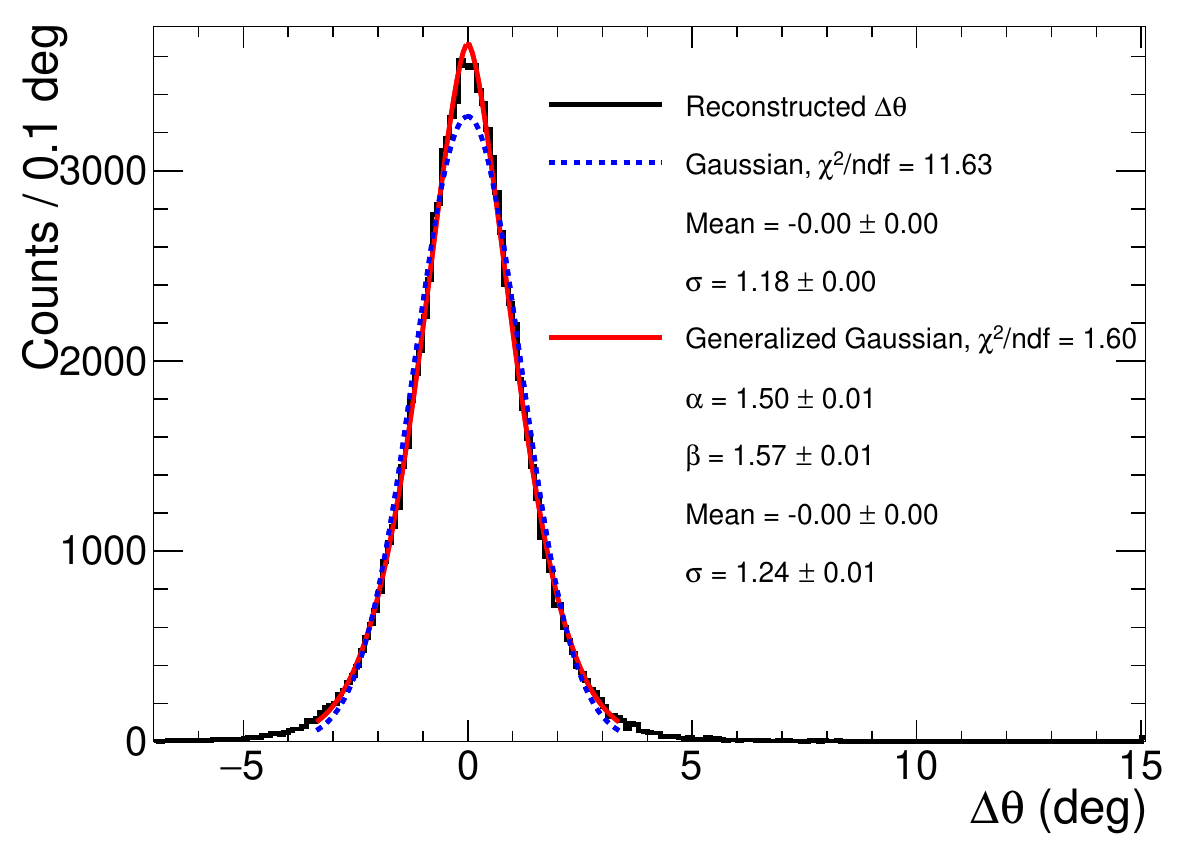}
\caption{ The relative incident angle for 1-GeV photons generated at $\theta=$~10$^{\circ}$. The distribution is fitted with the Gaussian and generalized Gaussian functions. The generalized Gaussian function describes the obtained distribution better.}
\label{fig:angle_10degree}
\end{figure}

Figure~\ref{fig:angle_10degree} represents the distribution of the relative incident angle ($\Delta\theta$) for 1-GeV photons generated at $\theta=$~10$^{\circ}$. The $\Delta\theta$ is calculated with the radial displacement of the reconstructed incident angle relative to the true incident direction. The central 98\% of the distribution is fitted with the Gaussian and generalized Gaussian (GG) functions~\cite{GGfun}. The GG function, also known as the generalized error distribution, shows better agreement with the obtained distribution, which is expressed as

\begin{eqnarray} 
f(x; \mu, \alpha, \beta) = \frac{\beta}{2 \alpha \Gamma(1/\beta)}e^{-(|x-\mu|/\alpha)^\beta},
\label{eqn:gg}
\end{eqnarray}
where $\mu$ is the mean value. The parameters $\alpha$ and $\beta$ determine the scale and shape of the distribution, respectively. The variance in the GG function is given by $\sigma^2 \equiv \alpha^2 \Gamma(3/\beta) / \Gamma(1/\beta)$ and the angular resolution is defined as $\sigma$. 

\begin{figure}[!hbt]
\centering
\includegraphics[width=0.79\textwidth]{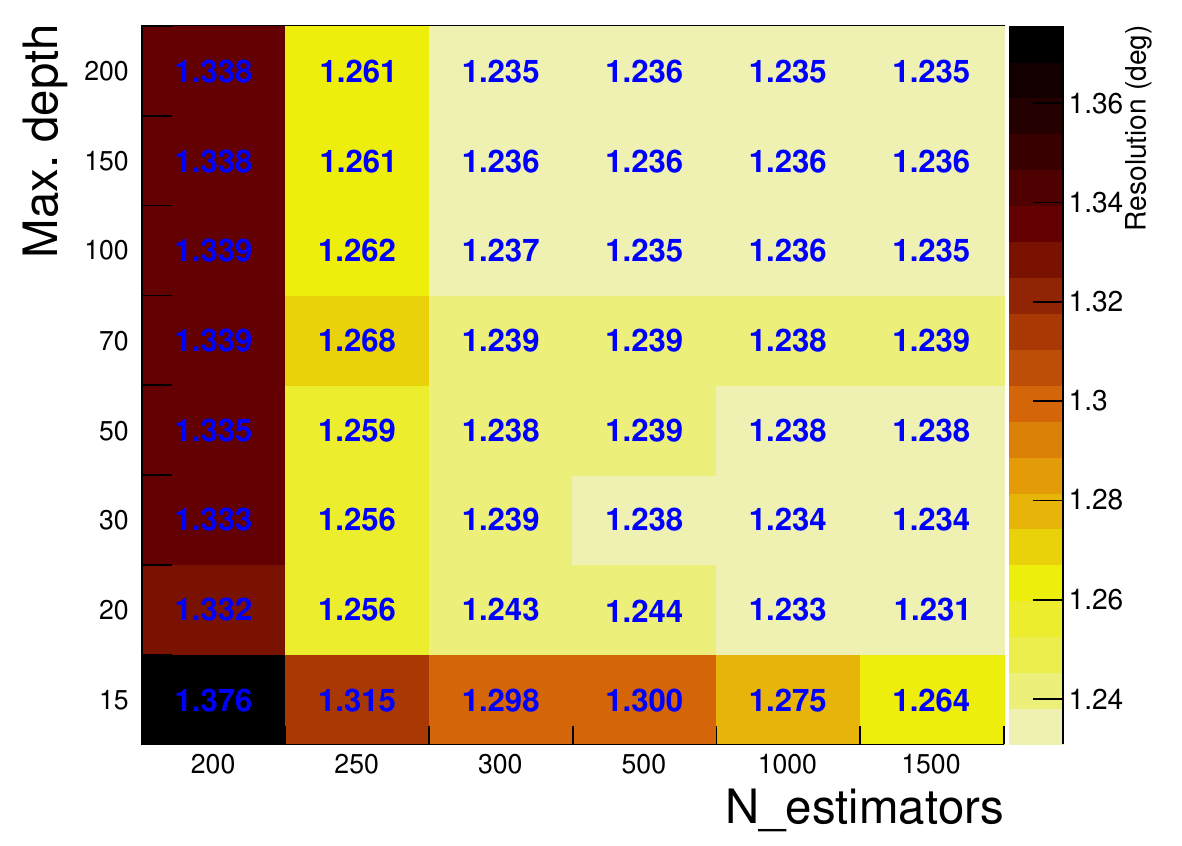}
\caption{Angular resolutions are displayed in terms of combination of \textit{N\_estimators} and \textit{Max. depth}. The uncertainty for each value is 0.01~deg. }
\label{fig:par_scan}
\end{figure}

Because the performance of the angle reconstruction of the XGB depends on the hyperparameters, we scan the angular resolution by changing them. Since there are correlations among the hyperparameters, the scanning process is performed in a five-dimensional space for all possible combinations. As an example, Figure~\ref{fig:par_scan} shows the test results for \textit{N\_estimators} and \textit{Max. depth}. The angular resolution improves when both \textit{Max. depth} and \textit{N\_estimators} increase. Since the uncertainty of the obtained angular resolution is estimated to be 0.01 degrees, many candidates for the hyperparameter set exist. Considering the CPU time for the training, the \textit{N\_estimators} and \textit{Max. depth} are set to 300 and 100, respectively. Similar processes are also performed for different hyperparameters, \textit{subsample}, \textit{learning rates}, and \textit{gamma}. Table~\ref{tab:XgbPar} lists the optimized set of hyperparameters that are used in further studies. During the optimization process, the fraction of the tail that is not well described by the GG function is approximately 2\%, and the fraction does not change for different sets of hyperparameters.

\begin{table}[hbt!]{\small
\centering
\caption{Hyperparameters of the XGB model}
\begin{tabular}{llcc}
\hline 
Parameter & Function & Default value & Used value \\ \hline 
\textit{N\_estimators} & The number of decision trees & N.A. & 300 \\  
\textit{Max. depth} & Possible maximum depth of tree structure & 6 & 100 \\ 
\textit{Subsample} & Fraction of total data used for a single decision & 1 & 0.8 \\ 
\textit{Learning rate} & Step length for calculation & 0.3 & 0.02 \\ 
\textit{Gamma} & Requirement on minimum loss function & 0 & 0 \\ 
\hline
\end{tabular}
\label{tab:XgbPar}
}\end{table}

\subsection{Performance of angular reconstruction}
\label{sec:perf}

\begin{figure}[!hbt]
\centering
\includegraphics[width=0.58\textwidth]{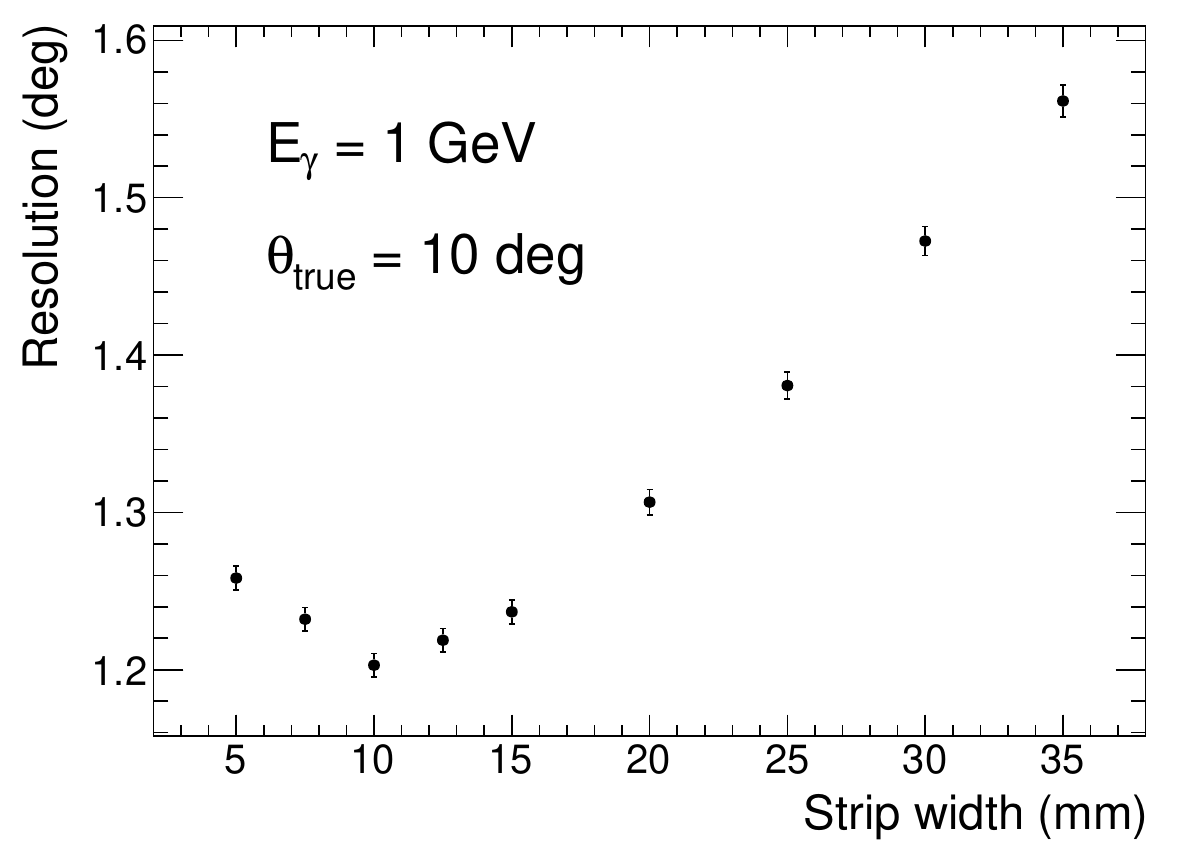}
\caption{ Angular resolutions are deduced in terms of the strip width for 1-GeV photons at $\theta=$~10$^{\circ}$ }
\label{fig:angle_reco_width}
\end{figure}
\subsubsection{Strip width}
To inspect the dependency of the angular resolution on the strip width, we deduce the angular resolution by varying the strip width from 5 to 35mm, as shown in Figure~\ref{fig:angle_reco_width}. The narrower the strip width, the larger the feature size. Consequently, the angle-reconstruction performance is better with 10-mm-wide strips than with 5-mm-wide strips. However, the wider strips hardly provide detailed information on the EM shower. The 15-mm-wide strips yield a good angular resolution of 1.24~$\pm$~0.01$^{\circ}$ and are selected for further study considering the actual construction of a prototype detector, although 10-mm-wide strips provide the best resolution.

\begin{figure}[!hbt]
\centering
\stackinset{c}{0.cm}{b}{-0.4cm}{(a)}{
\includegraphics[width=0.48\textwidth]{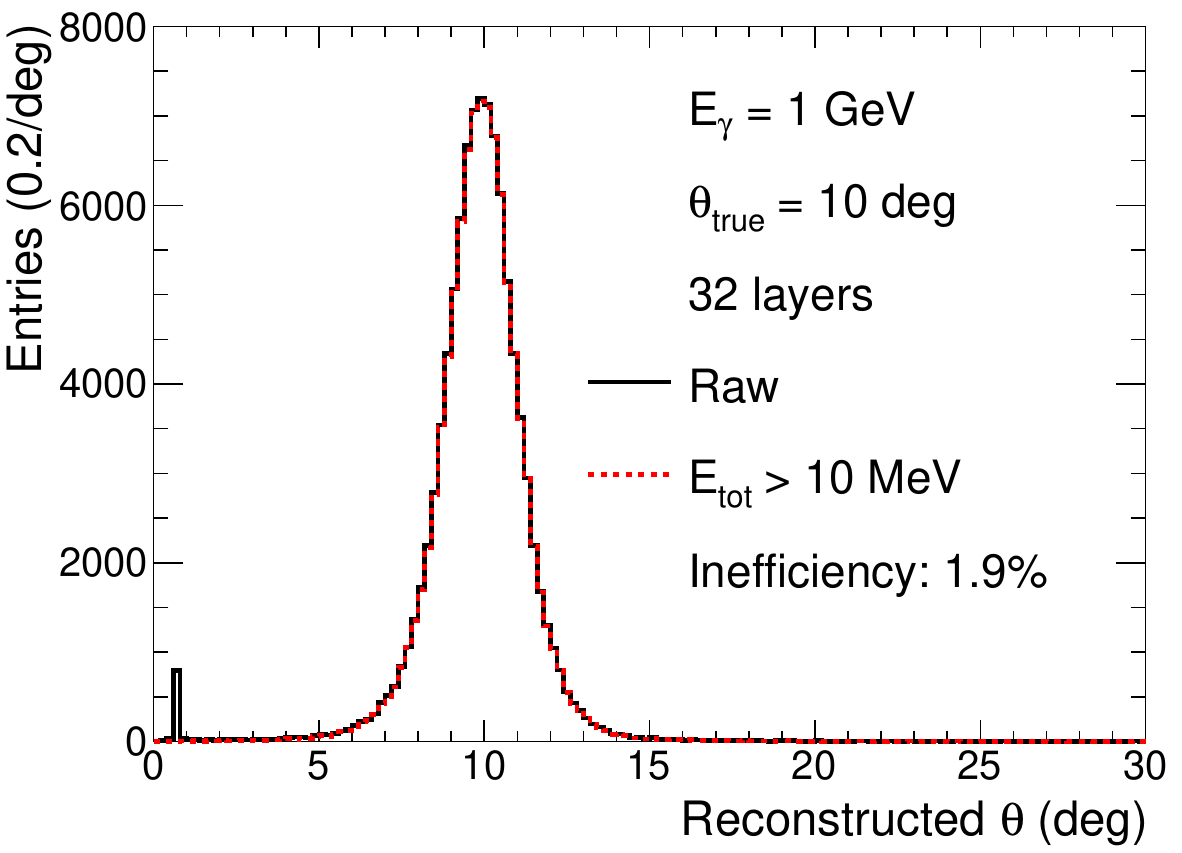}
}\stackinset{c}{0.cm}{b}{-0.4cm}{(b)}{ 
\includegraphics[width=0.48\textwidth]{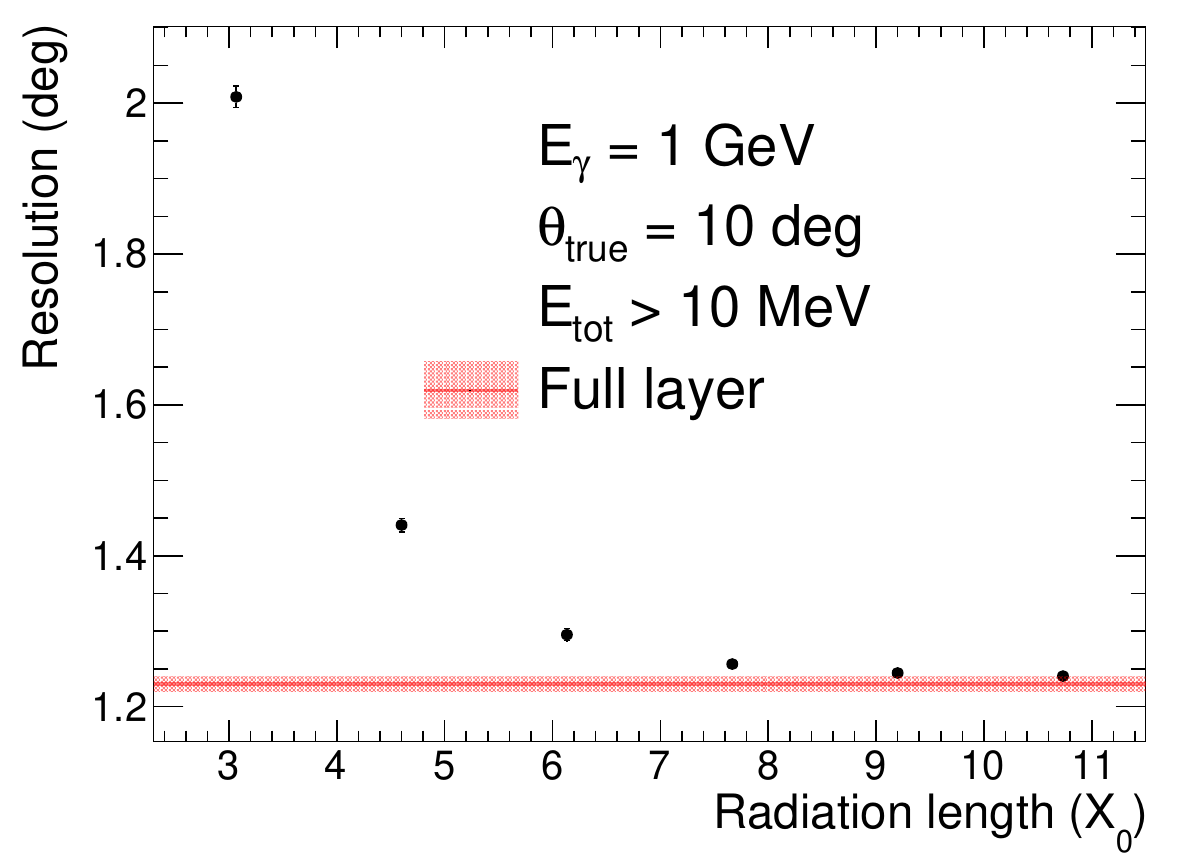}
}
\caption{ (a) Reconstructed polar angles ($\theta$) for 1-GeV photons at $\theta=$~10$^{\circ}$  with the front 32 layers of 15-mm-wide strips. The red histogram represents the angular distribution after requiring $E_{\rm{tot}}>$~10~MeV for the front layers. (b) Angular resolution as a function of the radiation length ($X_{0}$) with the front layers. The red line corresponds to the results using entire layers of 105.}
\label{fig:angle_reco_layer}
\end{figure}

\subsubsection{Number of layers}
We study the dependency of the angle reconstruction on the number of front layers. Figure~\ref{fig:angle_reco_layer} (a) shows the reconstructed angular distribution for 1-GeV photons by using only the front 32 layers, which correspond to 6.2$X_{0}$. A fraction of the photons fails to be reconstructed because of insufficient information gathering caused by deeply penetrating photons without shower generation, where the fraction is estimated to be 1.9\%. The failed events are represented as a delta function near 0. Such events are removed by requiring the total energy deposit at the front layers to be higher than 10~MeV, which is 1\% of the incident energy. The angular resolution with the front 32 layers is estimated to be 1.30~$\pm$~0.01$^{\circ}$. The evolution of angular resolution degrades with the reduction of the number of layers used.

\begin{figure}[!hbt]
\centering
\stackinset{c}{0.cm}{b}{-0.4cm}{(a)}{
\includegraphics[width=0.44\textwidth]{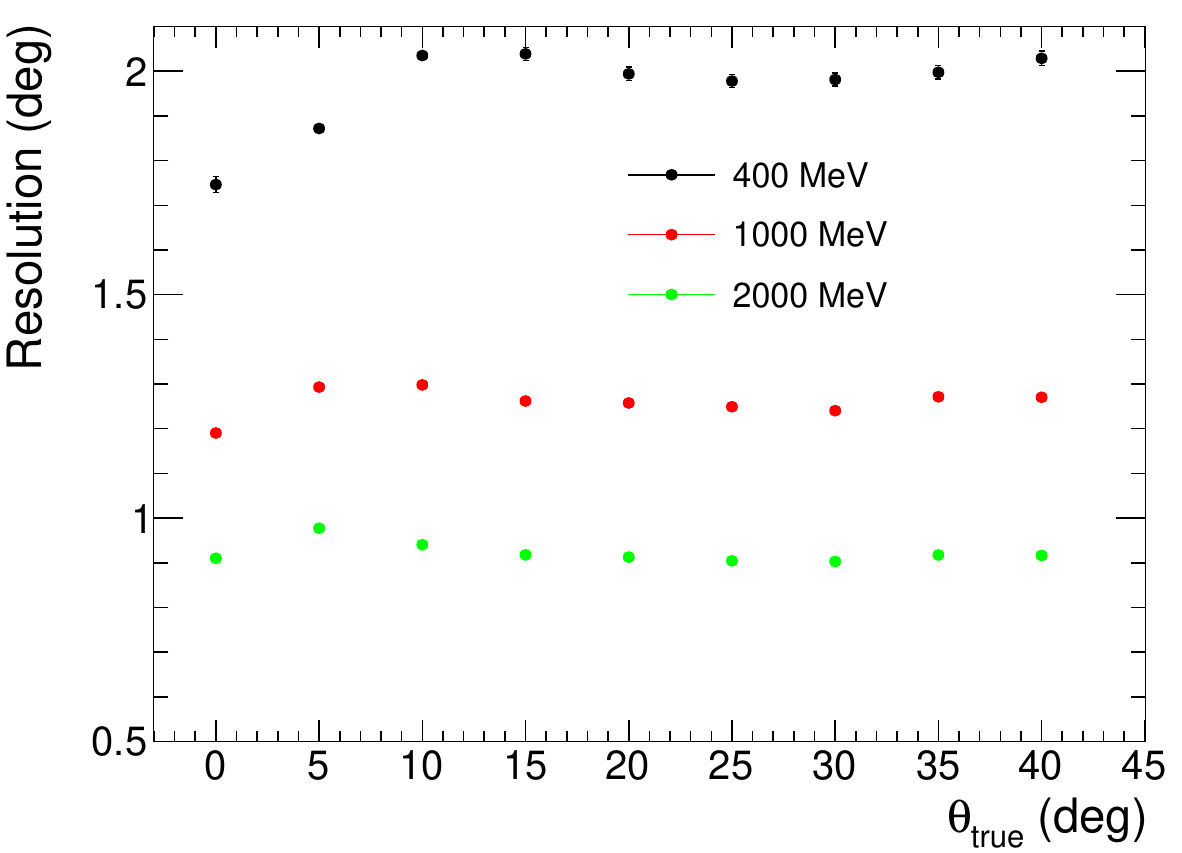}
}\stackinset{c}{0.cm}{b}{-0.4cm}{(b)}{
\includegraphics[width=0.44\textwidth]{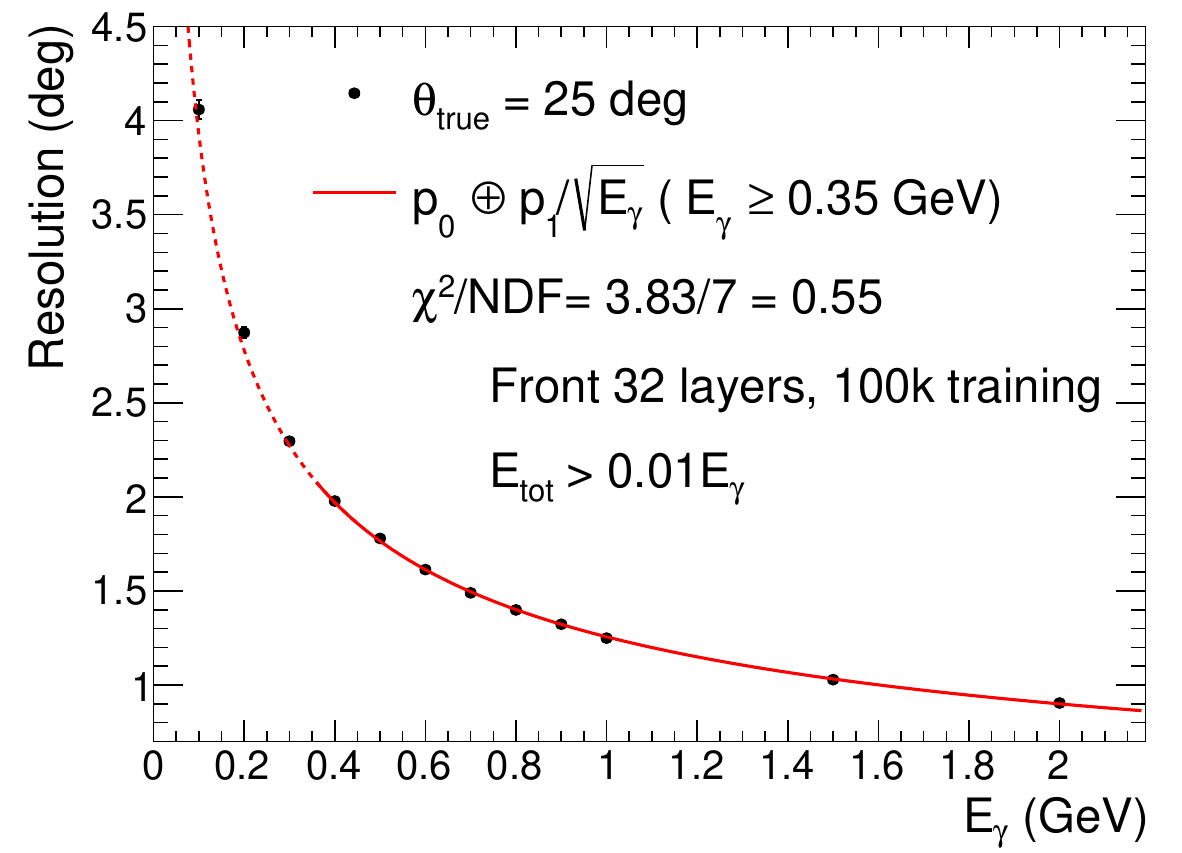}
}
\caption{ (a) Angular resolution as a function of the incident angle for different incident energies. (b) Angular resolution as a function of the incident energy for $\theta=$~25$^{\circ}$; it is fitted to the function $p_{0} \oplus p_{1}/\sqrt{E_{\gamma}(\mathrm{GeV})}$. All resolutions are estimated with the front 32 layers. }
\label{fig:angle_reco_dep_gr}
\end{figure}

\subsubsection{Incident energy}
Figure~\ref{fig:angle_reco_dep_gr} (a) shows the angular resolution as a function of incident angle for different incident energies ($E_{\gamma}$). Note that the front 32 layers are used for the angle reconstruction. Since the effective radiation length is different according to the incident angles, the inefficiency of the reconstruction is changed from 2\% at zero degrees to 0.5\% at 40 degrees. The angular resolution does not depend on the incident angle for high incident energies. However, the angular resolution changes significantly for low incident energies at small incident angles. Figure~\ref{fig:angle_reco_dep_gr} (b) shows the angular resolution as a function of the incident photon energy at $\theta=$~25$^{\circ}$. The angular resolution is well fitted by the function $p_{0} \oplus p_{1}/\sqrt{E_{\gamma}(\mathrm{GeV})}$, where $p_{0}$ and $p_{1}$ represent the energy-independent and energy-dependent contributions, respectively, added in quadrature.

\begin{figure}[!hbt]
\centering
\includegraphics[width=0.44\textwidth]{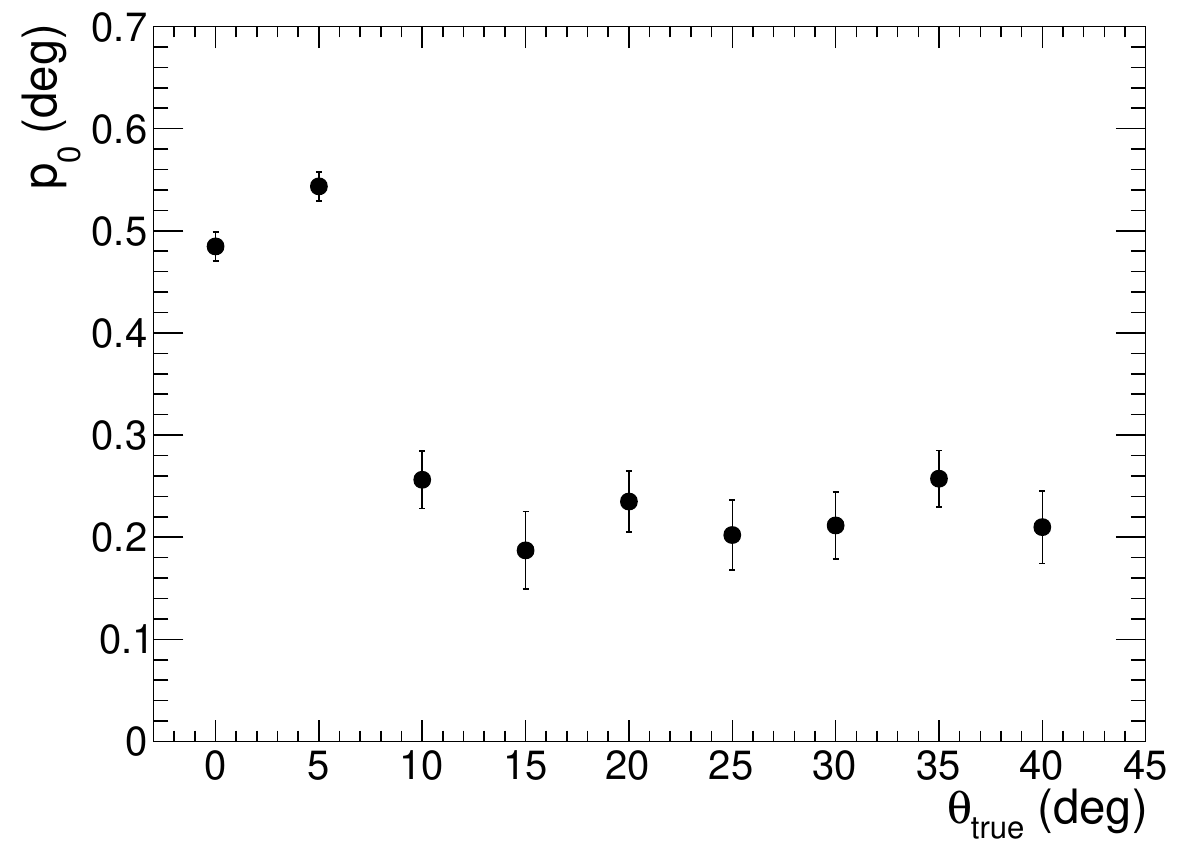}
\includegraphics[width=0.44\textwidth]{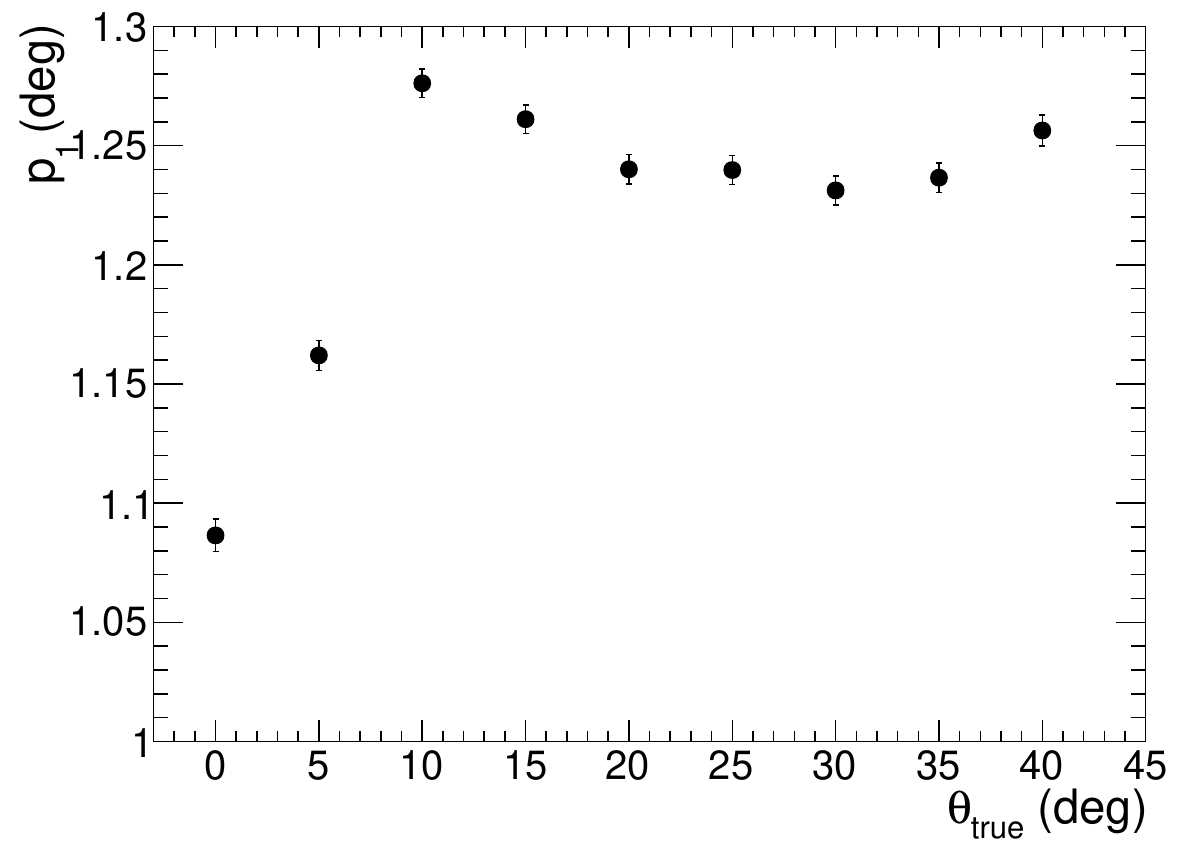}
\caption{ $p_{0}$ and $p_{1}$ as a function of the incident angle }
\label{fig:res_edep}
\end{figure}

Figure~\ref{fig:res_edep} shows the estimated $p_{0}$ and $p_{1}$ as a function of the incident angle. The averages of $p_{0}$ and $p_{1}$ are estimated to be 0.238 $\pm$ 0.012$^{\circ}$ and 1.248 $\pm$ 0.002$^{\circ}$, respectively, for $\theta>$~10$^{\circ}$. In the case of $\theta$ smaller than 10 degrees, the angular resolution shows different dependency on the incident energy, which results in smaller $p_{1}$ and larger $p_{0}$ compared to the larger angles.

\begin{figure}[!hbt]
\centering
\includegraphics[width=0.6\textwidth]{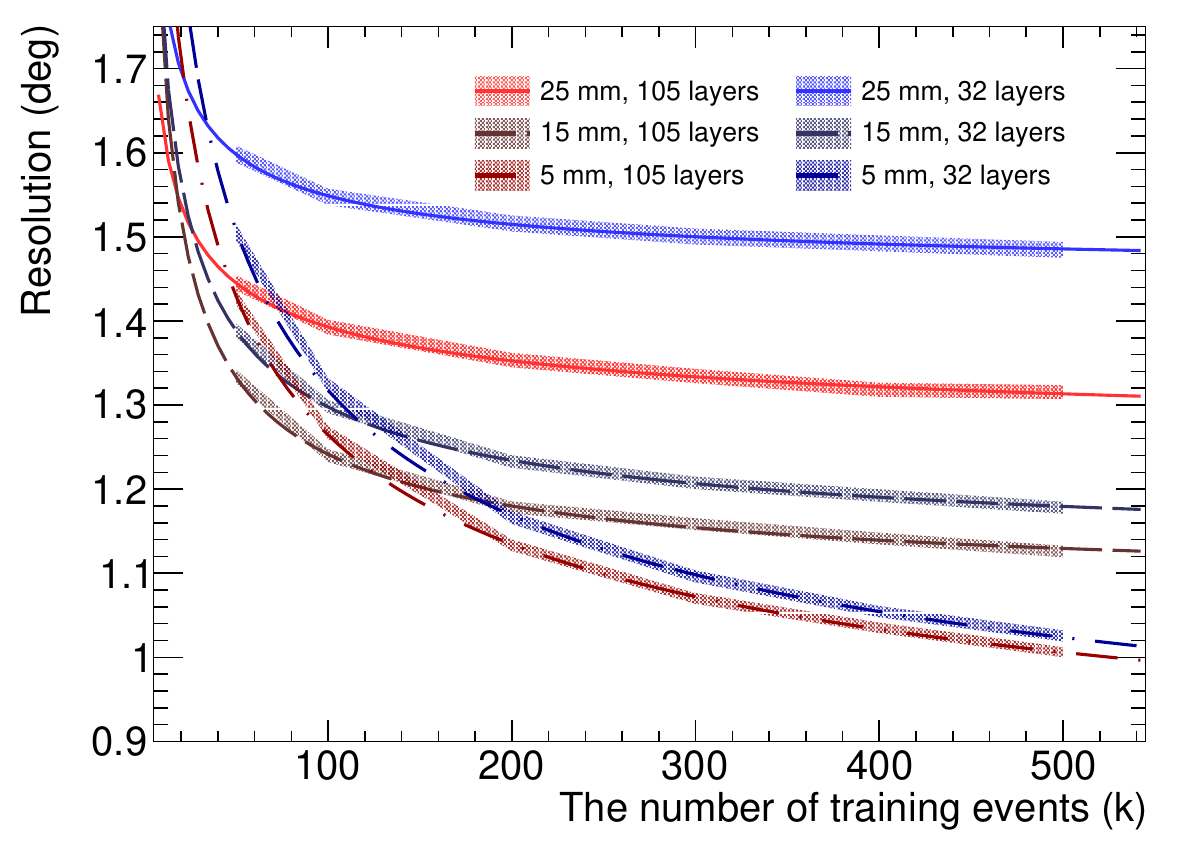}
\caption{ The angular resolution for 1-GeV photons at $\theta=$~10$^{\circ}$ as a function of the number of training samples with different strip widths obtained using the first 32 layers or the full layers. The spectra are fitted with $\sqrt{I^{2} + (D/N_{\mathrm{sample}})^{k}}$. }
\label{fig:multi-parameter}
\end{figure}

\subsubsection{Training sample}
Figure~\ref{fig:multi-parameter} shows the angular resolution for different numbers of training samples and detector configurations. The width of each curve represents the statistical uncertainty. All configurations show improvement in the angular resolution with the increasing number of training samples. With an increase in the number of training events, the angular resolution improves most rapidly for the configuration with 5-mm-wide scintillator strips. In other words, the required number of training events should be large enough to provide correct answer from the correlations between many features.

The spectra are fitted with the empirical function of $\sqrt{I^{2} + (D/N_{\mathrm{sample}})^{k}}$, where $I$ denotes the intrinsic resolution and $D$ and $k$ describe the decreasing trend. The numbers of training samples reaching the angular resolution of 1.1$I$ with 25-, 15-, and 5-mm-wide strips are estimated to be 58k, 280k, and 7.6M, respectively, indicating that the number of training samples are insufficient especially for 5-mm-wide strips.

\section{Summary}
\label{sec:sum}

A sampling calorimeter for the incident angle measurement of photons will play an essential role in the next stage of the KOTO experiment, which can distinguish $\pi^0$ decays at the off- from in-beam axis. The machine-learning approach (XGBoost) analyzing the energy-weighted shower shapes enables us to proceed with detailed optimization of the detector to fulfill the experimental goal. The angular resolution depends on the values of hyperparameters of the XGBoost decided by the user.

The width of strips to record the energy deposit of shower particles is an essential parameter of the angular resolution, which should be compromised between the required angular resolution and construction costs related to the number of read-out channels. In addition, the angular resolution is also determined by the training accuracy of the XGBoost, which requires a different number of training data according to the detector configuration. When we test a detector configuration with many channels (i.e., narrow strip), we should also prepare a large number of data for the training.

For the toy sampling calorimeter with 15-mm-wide strips and 32 layers, the angular resolution was obtained as 1.30~$\pm$~0.01$^{\circ}$ in machine learning with $10^{5}$ training samples. Energy dependence of the angular resolution can be expressed as $p_{0} \oplus p_{1}/\sqrt{E_{\gamma}}$. For the angular range of $10^{\circ}<\theta<40^{\circ}$, the $p_{0}$ and $p_{1}$ remain unchanged as $p_{0}$ = 0.238 $\pm$ 0.012$^{\circ}$ and $p_{1}$ = 1.248 $\pm$ 0.002$^{\circ}$, respectively.

\label{sec:con}


\section*{Acknowledgment}
This work was supported by the National Research Foundation of Korea (NRF) 
(Grants No. 2019R1A2C1084552, 2022R1A5A1030700, and 2020R1A3B2079993) and JSPS KAKENHI Grant No. JP21H01118.

\end{document}